\documentclass[aps,prl,twocolumn,superscriptaddress,preprintnumbers,amsmath,amssymb]{revtex4}

\usepackage{graphicx} 
\usepackage{dcolumn}  
\usepackage{rotating}
\usepackage{booktabs}
\usepackage{SIunits} 
\usepackage{graphicx}
\usepackage{dcolumn}
\usepackage{diagbox}
\usepackage{bm}
\usepackage{color}
\usepackage[english]{babel}
\usepackage{titlesec}
\usepackage{multirow}
\usepackage[colorlinks,linkcolor=blue,urlcolor=blue,anchorcolor=green,citecolor=blue,breaklinks=true]{hyperref}
\usepackage{epstopdf}
\usepackage{enumitem}
\usepackage{hhline}
\usepackage{array}
\usepackage{threeparttable}
\usepackage{ulem}
\usepackage{url}
\usepackage{hyperref}
\usepackage{orcidlink} 

\usepackage[hyphenbreaks]{breakurl}

\newcommand{\BR}{{\cal B}}

\newcommand{\EE}{e^+e^-}

\begin{document}
	\hyphenpenalty=10000
	\tolerance=1000
	\vspace*{-3\baselineskip}
	
\noaffiliation
\author{Y.~B.~Li\,\orcidlink{0000-0002-9909-2851}} 
\author{C.~P.~Shen\,\orcidlink{0000-0002-9012-4618}} 
\author{I.~Adachi\,\orcidlink{0000-0003-2287-0173}} 
\author{H.~Aihara\,\orcidlink{0000-0002-1907-5964}} 
\author{D.~M.~Asner\,\orcidlink{0000-0002-1586-5790}} 
\author{H.~Atmacan\,\orcidlink{0000-0003-2435-501X}} 
\author{T.~Aushev\,\orcidlink{0000-0002-6347-7055}} 
\author{R.~Ayad\,\orcidlink{0000-0003-3466-9290}} 
\author{V.~Babu\,\orcidlink{0000-0003-0419-6912}} 
\author{S.~Bahinipati\,\orcidlink{0000-0002-3744-5332}} 
% ...
\collaboration{The Belle Collaboration}
	\title{\boldmath Evidence of a new excited charmed baryon decaying to $\Sigma_{c}(2455)^{0,++} \pi^{\pm}$}

	\begin{abstract}
		
		We present the study of $\bar{B}^{0} \to \Sigma_{c}(2455)^{0,++} \pi^{\pm} \bar{p}$ decays based on $772\times10^{6}$ $B\bar{B}$ events collected with the Belle detector at the KEKB asymmetric-energy $e^+e^-$ collider. The $\Sigma_{c}(2455)^{0,++} $ candidates are reconstructed via their decay to $\Lambda_{c}^{+} \pi^{\mp}$ and $\Lambda_{c}^{+}$ decays to $pK^{-}\pi^{+},~pK_{S}^{0},$ and $\Lambda\pi^{+}$ final states. The corresponding branching fractions are measured to be $\BR(\bar{B}^{0} \to \Sigma_{c}(2455)^{0} \pi^{+} \bar{p}) = (1.09 \pm 0.06 \pm 0.07)\times10^{-4}$ and $\BR(\bar{B}^{0} \to \Sigma_{c}(2455)^{++} \pi^{-} \bar{p}) = (1.84\pm 0.11 \pm 0.12)\times10^{-4}$, which are consistent with the world average values with improved precision. A new structure is found in the $M_{\Sigma_{c}(2455)^{0,++}\pi^{\pm}}$ spectrum with a significance of $4.2\sigma$ including systematic uncertainty. The structure is possibly an excited $\Lambda_{c}^{+}$ and is tentatively named $\Lambda_{c}(2910)^{+}$. Its mass and width are measured to be $(2913.8 \pm 5.6 \pm 3.8)$ MeV/$c^{2}$ and $(51.8\pm20.0 \pm 18.8)$ MeV, respectively. The products of branching fractions for the $\Lambda_{c}(2910)^{+}$ are measured to be $\BR(\bar{B}^{0} \to \Lambda_{c}(2910)^{+}\bar{p})\times\BR(\Lambda_{c}(2910)^{+} \to \Sigma_{c}(2455)^{0}\pi^{+}) = (9.5 \pm 3.6 \pm 1.6)\times10^{-6}$ and $\BR(\bar{B}^{0} \to \Lambda_{c}(2910)^{+}\bar{p})\times\BR(\Lambda_{c}(2910)^{+} \to \Sigma_{c}(2455)^{++}\pi^{-}) = (1.24 \pm 0.35 \pm 0.10)\times10^{-5}$.
		Here, the first and second uncertainties are statistical and systematic, respectively.
		
	\end{abstract}

	\maketitle
	
	\tighten
	
	{\renewcommand{\thefootnote}{\fnsymbol{footnote}}}
	\setcounter{footnote}{0}


	Due to the numerous degrees of freedom of the internal structure of charmed baryons, their spectroscopy provides an excellent laboratory for studying the dynamics of light quarks in the environment of a heavy quark and testing heavy-quark symmetry and chiral symmetry of light quarks~\cite{Yan:1992gz,Burdman:1992gh,Wise:1992hn}. Although many excited charmed baryons have been discovered by the BaBar, Belle, CLEO, and LHCb in the past two decades~\cite{PDG}, there are still missing states in the predicted spectra~\cite{cisc} and properties of many known particles are still poorly understood~\cite{PDG}.
	
	Currently, there is no unified phenomenological model that can fully describe the charmed baryon sector. Theoretically, the mass spectrum of excited charmed baryons has been studied with numerous approaches, such as a Quantum Chromodynamics (QCD) based quark model~\cite{Copley:1979wj}, the QCD sum rule~\cite{Wang1, Wang2, Ye, sundu, Chen3}, Reggie phenomenology~\cite{Guo}, a relativistic quark potential model~\cite{Capstick:1986ter}, quark potential models~\cite{Majethiya,Migura, Garcilaz, Ebert1, Ebert2}, the relativistic flux tube models~\cite{Chen1, Chen2}, the coupled channel model~\cite{Romanets}, the constituent quark models~\cite{kai,bing,yoshida}, and lattice QCD~\cite{Padmanath1, Padmanath2}. More experimental measurements are required to validate these theoretical models.
	
	Among the observed excited $\Lambda_{c}^{+}$ family, the highest state $\Lambda_{c}(2940)^{+}$ presents many mysteries. It was discovered by BaBar via its decay to $D^{0}p$~\cite{BaBar:2006itc}, and confirmed by LHCb~\cite{LHCb:2017jym}, and its decay to $\Sigma_{c}(2455)^{0,++}\pi^{\pm}$ was observed by Belle~\cite{Belle:2006xni}. Though the quantum number $J^{P} = \frac{3}{2}^{-}$ is favored for $\Lambda_{c}(2940)^{+}$ according to the LHCb measurement, other spin-parity assignments are also proposed~\cite{Cheng:2015iom,cisc}. Besides that, the mass of $\Lambda_{c}(2940)^{+}$ is lower than the expected $\Lambda_{c}(\frac{3}{2}^{-},2P)$ state in the quark models~\cite{Capstick:1986ter,Ebert2,Chen1,bing}, in which its mass is expected to be above 3 GeV/$c^{2}$ and the mass of the undiscovered $\Lambda_{c}(\frac{1}{2}^{-},2P)$ state is slightly lighter than that of $\Lambda_{c}(\frac{3}{2}^{-},2P)$ by not more than 25 MeV/$c^{2}$. Such a low-mass puzzle for $\Lambda_{c}(2940)^{+}$ can be explained by introducing the $D^{*}N$ channel contribution~\cite{Luo:2019qkm}, while the mass of $\Lambda_{c}(\frac{1}{2}^{-},2P)$ state is higher than that of $\Lambda_{c}(\frac{3}{2}^{-},2P)$ by around 40 MeV/$c^{2}$ in this scenario, which leads to an interesting mass inversion. Thus, it is important to verify the quantum number of $\Lambda_{c}(2940)^{+}$ or search for other candidates of $\Lambda_{c}(2P)$.
	
	Compared to the previous inclusive analyses~\cite{BaBar:2006itc,LHCb:2017jym,Belle:2006xni}, the study of $\Lambda_{c}(2P)$ can be performed in $\bar{B}^{0} \to \Lambda_{c}^{+}(2P) (\to \Sigma_{c}(2455)^{0,++} \pi^{\pm}) \bar{p}$ exclusive decays, which can constrain the spin and parity of the possible excited $\Lambda_{c}^{+}(2P)$ and provide a simpler background environment. The $\bar{B}^{0} \to \Sigma_{c}(2455)^{0,++} \pi^{\pm} \bar{p}$ decays have been previously studied by CLEO~\cite{cleo_old}, Belle~\cite{belle_old,belle_old2}, and BaBar~\cite{babar_old} based on 9.17 fb$^{-1}$, 357 fb$^{-1}$ and 426 fb$^{-1}$ $\Upsilon(4S)$ data samples, respectively, with $\Lambda_{c}^{+}$ reconstructed via the $pK^{-}\pi^{+}$ mode. The average branching fractions are $ \BR(\bar{B}^{0} \to \Sigma_{c}(2455)^{0} \pi^{+} \bar{p}) = (1.08 \pm 0.16) \times10^{-4}$ and $\BR(\bar{B}^{0} \to \Sigma_{c}(2455)^{++} \pi^{-} \bar{p}) = (1.88 \pm 0.24 )\times10^{-4}$. The invariant mass spectra of $M_{\Sigma_{c}(2455)^{0,++} \pi^{\pm}}$ and $M_{\bar{p}\pi^{\pm}}$ are found to be inconsistent with phase-space distributions. In particular, Belle's analysis in Ref.~\cite{belle_old2} suggested that there could be a structure or overlap of several known excited $\Lambda_{c}^{+}$ near the threshold of the $M_{\Sigma_c(2455)^{0}\pi^{+}}$ spectrum, which needs further study.
	
	In this Letter, we present a study of the $\bar{B}^{0} \to \Sigma_{c}(2455)^{0,++} \pi^{\pm} \bar{p}$ decays~\cite{charge-conjugate} and study the possible resonance in the $M_{\Sigma_{c}(2455)^{0,++} \pi^{\pm}}$ spectrum using the full data sample of 711 fb$^{-1}$
	collected at the $\Upsilon(4S)$ resonance by the Belle detector~\cite{Belle} at the
	KEKB asymmetric energy electron-positron collider~\cite{KEKB}.
	Simulated signal events with $\bar{B}^{0}$ meson decays are generated using {\sc EvtGen}~\cite{evtgen}. These events are processed by a
	detector simulation based on
	{\sc GEANT3}~\cite{geant3}. The generic Monte Carlo (MC) samples of $\Upsilon(4S)\to B \bar{B}$ ($B=B^+$ or $B^0$) and $e^+e^- \to q \bar{q}$
	($q=u,~d,~s,~c$) events at $\sqrt{s}=10.58$ GeV are used to
	check the backgrounds~\cite{zhouxy_topo}, corresponding to five times the integrated luminosity of the data.
	
	For charged track identification, information from different detector subsystems is combined to form the likelihood $\mathcal{L}_{i}$ for species $i$, where $i=\pi$,~$K$, or $p$~\cite{pid}. Except for the charged tracks
	from $\Lambda \to p \pi^{-}$ and $K_{S}^{0} \to \pi^{+} \pi^{-}$
	decays, a track with a likelihood ratio $\mathcal{R}_K^{\pi} =
	\mathcal{L}_K/(\mathcal{L}_K + \mathcal{L}_\pi)> 0.6~(<0.4)$ is
	identified as a kaon (pion)~\cite{pid}. With this selection, the kaon (pion)
	identification efficiency is about 93\% (97\%). A track with
	$\mathcal{R}^\pi_{p} =
	\mathcal{L}_{p}/(\mathcal{L}_{p}+\mathcal{L}_\pi)
	> 0.6$ and $\mathcal{R}^K_{p} =
	\mathcal{L}_{p}/(\mathcal{L}_{p}+\mathcal{L}_K) >
	0.6$ is identified as a proton with an efficiency above
	90\%. The $K_{S}^{0}$ and $\Lambda$ candidates are reconstructed from pairs of
	oppositely charged tracks, treated as $\pi^{+} \pi^{-}$ and $ p \pi^-$, with the similar method used in Ref.~\cite{Xic2930}. The $p \pi^-$ invariant mass should be within 3.5 MeV/$c^2$ ($\sim$$3\sigma$, where $\sigma$ denotes the mass resolution) of the $\Lambda$ nominal mass~\cite{PDG}. The $\Sigma_{c}(2455)^{0,++}$ candidates are reconstructed via their decay to $\Lambda_{c}^{+} \pi^{\mp}$, while the $\Lambda_c^+$ are reconstructed with the $\Lambda_c^+\to p K^- \pi^+$,
	$pK_{S}^{0}$, and $\Lambda \pi^{+}$. The mass windows of $\Sigma_{c}(2455)^{0,++}$ and $\Lambda_{c}^{+}$ are within 10 MeV/$c^{2}$ and 14 MeV/$c^{2}$ of their nominal masses~\cite{PDG}, respectively, which retain more than 94\% of the signal events. About 8\% of events have multiple candidates that are all used for further analysis.
	
\begin{figure}[htbp]
		\begin{center}
			\includegraphics[width=7cm]{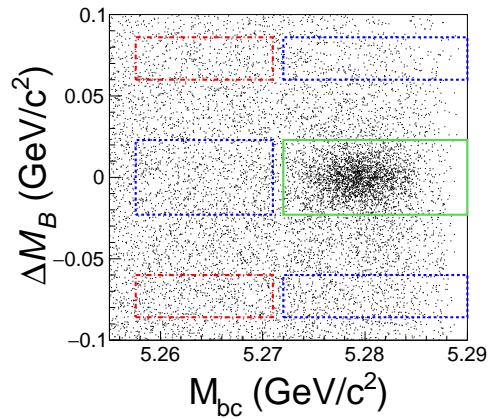}	
			\caption{The scatter distribution of $\Delta M_{B}$ versus $M_{\rm bc}$ from data. The blue and red boxes are the $B$ sideband regions described in the text. The green box indicates the signal region. \label{mbcdm} }
		\end{center}
	\end{figure}

	Figure~\ref{mbcdm} shows the scatter plot of $\Delta M_{B}$ versus $M_{\rm bc}$ of the selected $\bar{B}^{0} \to \Sigma_{c}(2455)^{0,++} \pi^{\pm} \bar{p}$ candidates from data after applying the selection criteria above. The $M_{\rm bc}$ is defined as $\sqrt{E_{\rm beam}^{2}/c^2 - (\sum \vec{p}_{i})^2}/c$, where $E_{\rm beam}$ and $\vec{p}_{i}$ are the beam energy and the three-momenta of the $\bar{B}^{0}$-meson decay products in the center-of-mass system of the $\EE$ collision. The $\Delta M_{B}$ is defined as $M_{B} - m_{B}$, where $M_{B}$ is the invariant mass of the $\bar{B}^{0}$ candidate and $m_{B}$ is the nominal $\bar{B}^{0}$-meson mass~\cite{PDG}. The $\bar{B}^{0}$ signal region is $|\Delta M_{B}| < 0.023$ GeV/$c^{2}$ ($\sim 2.5 \sigma$) and $M_{\rm bc} > 5.272$ GeV/$c^{2}$ ($\sim 2.5 \sigma$), which is illustrated by the green box in Fig.~\ref{mbcdm}.
	
\begin{figure}[htbp]
		\begin{center}
			\includegraphics[width=4.2cm]{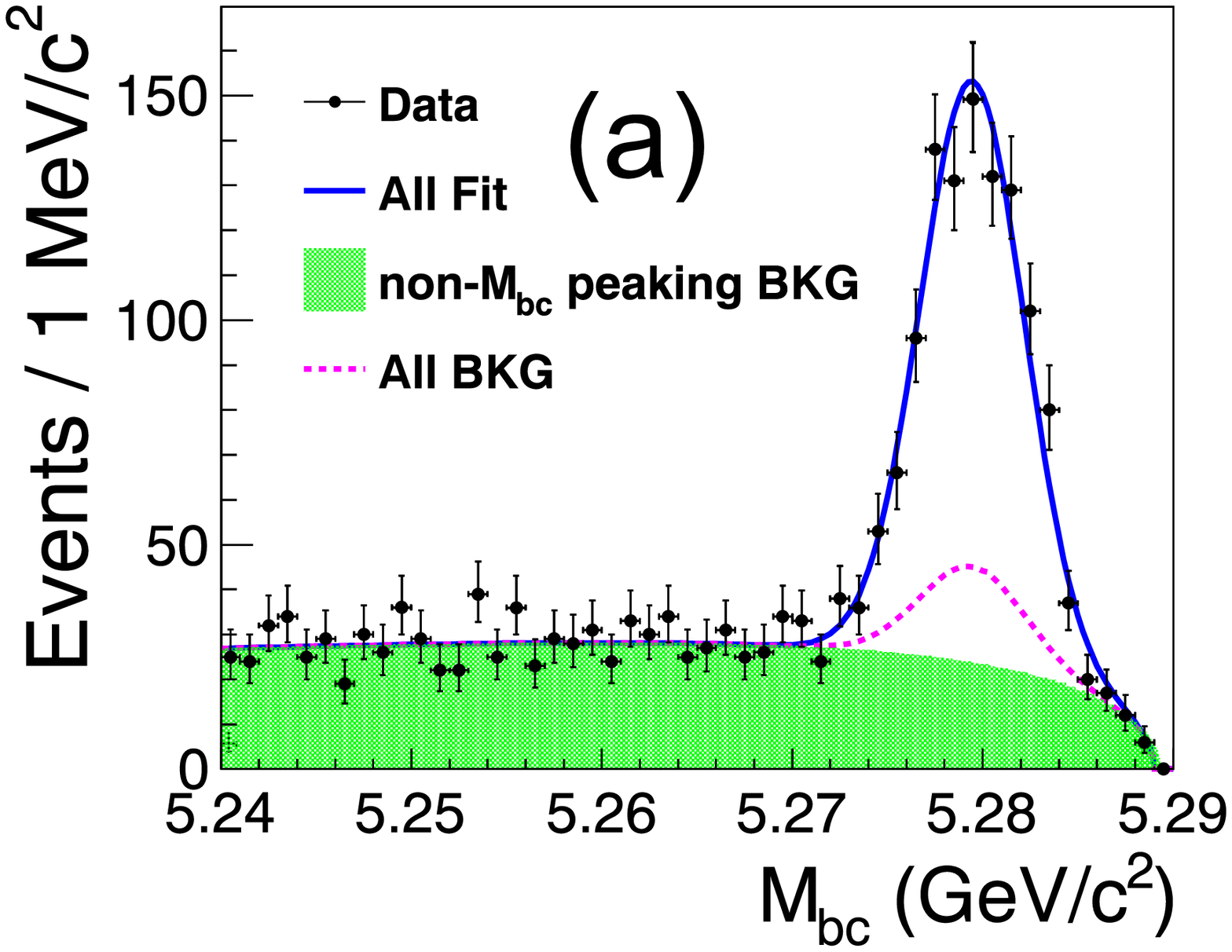}
			\includegraphics[width=4.2cm]{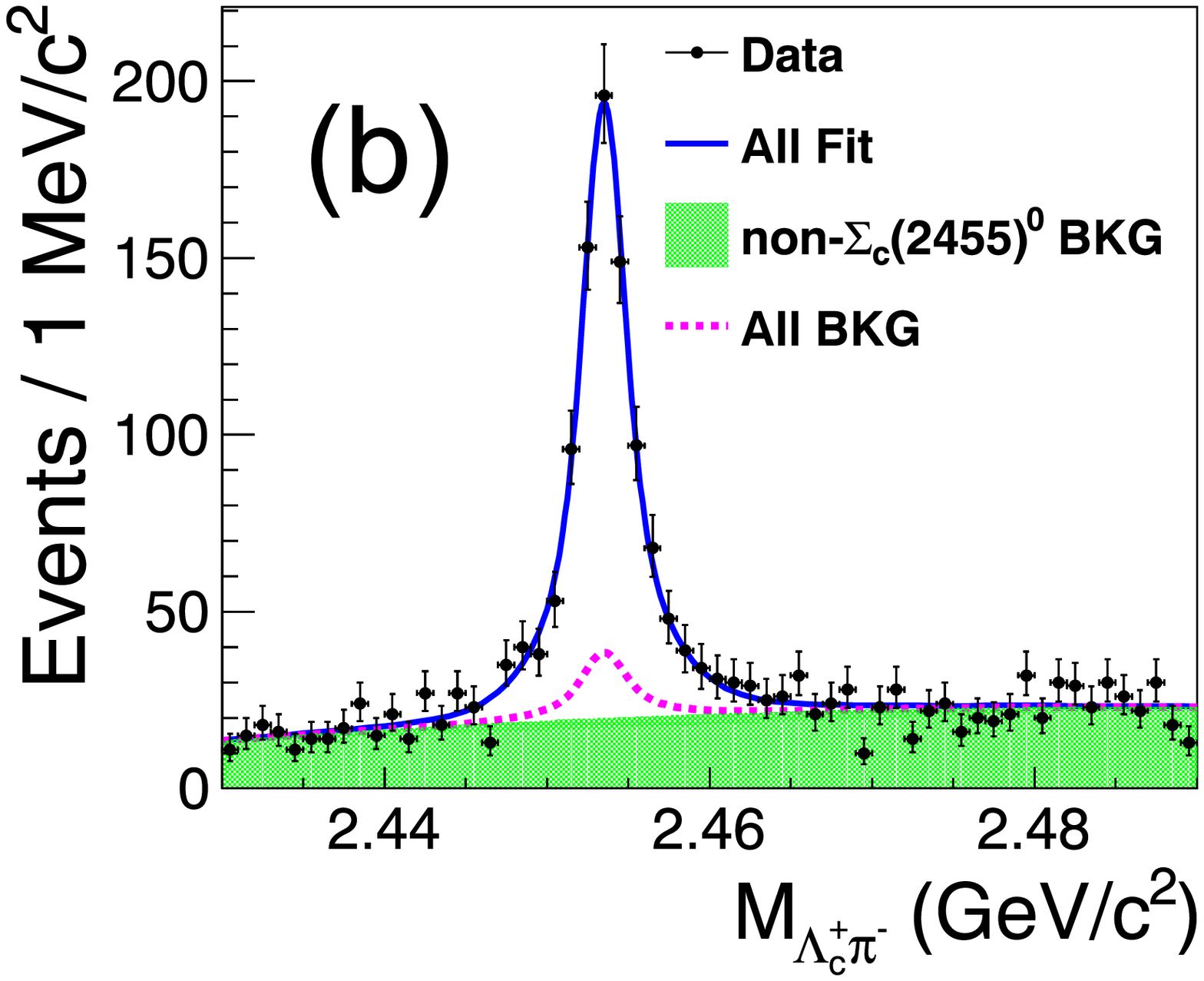}
			
			\includegraphics[width=4.2cm]{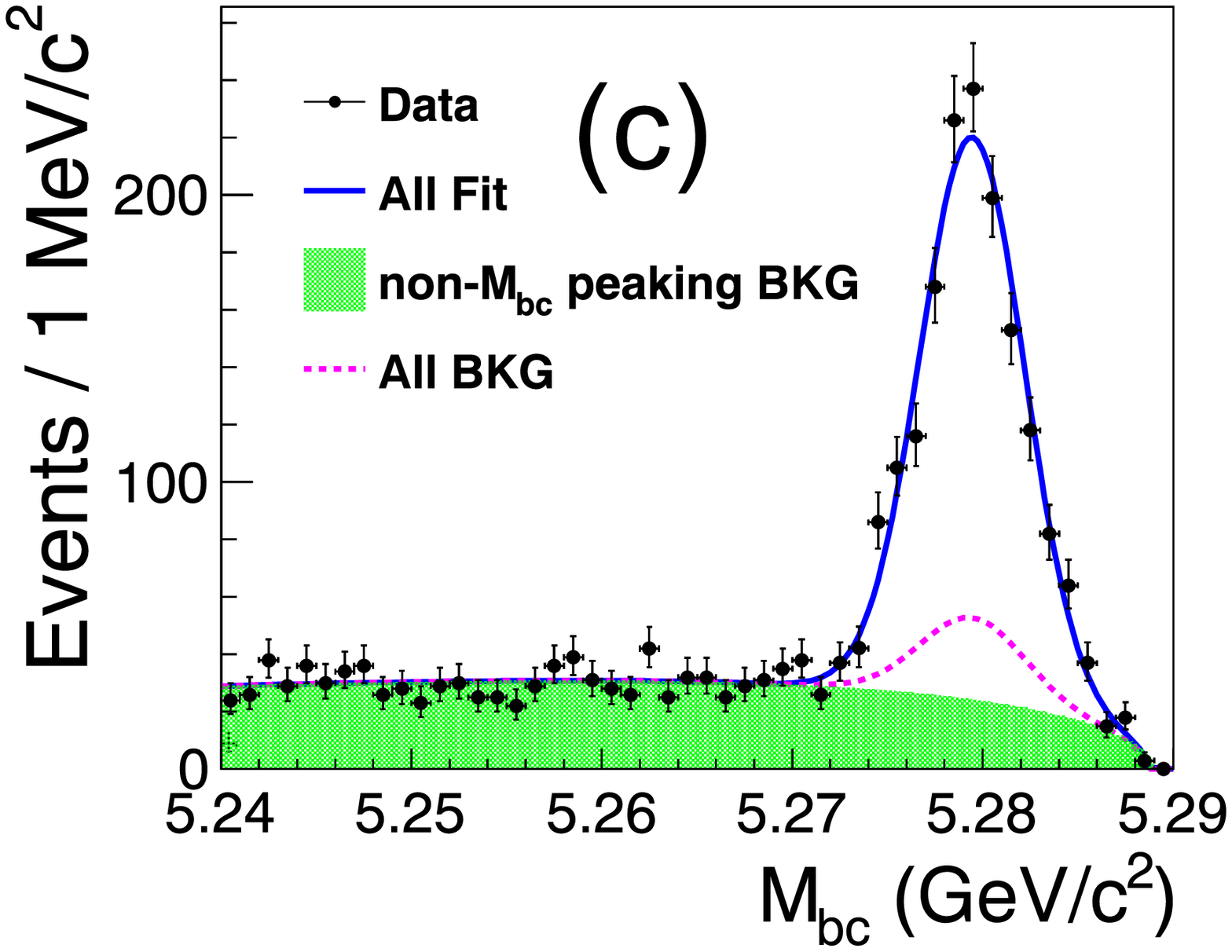}
			\includegraphics[width=4.2cm]{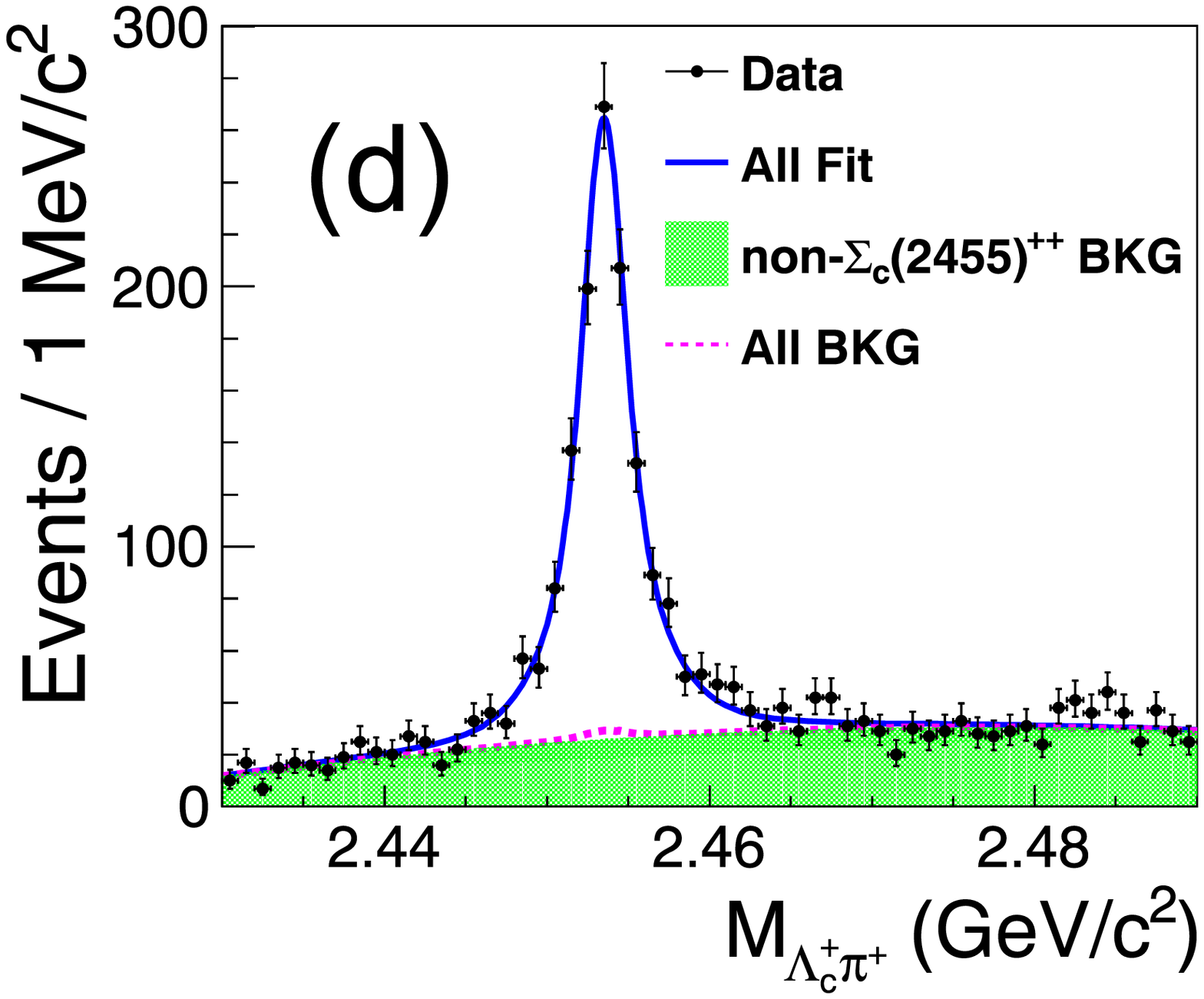}
			
			\caption{\label{2dfitB} The projections of (a) $M_{\rm bc}$ and (b) $M_{\Lambda_{c}^{+}\pi^{-}}$ of the 2D fit to the selected $\bar{B}^{0} \to \Sigma_{c}(2455)^{0}\pi^{+} \bar{p}$ candidates, and the projections of (c) $M_{\rm bc}$ and (d) $M_{\Lambda_{c}^{+}\pi^{+}}$ of the 2D fit to the selected $\bar{B}^{0} \to \Sigma_{c}(2455)^{++}\pi^{-} \bar{p}$ candidates. The dots with error bar are from data; the blue solid curves are the best fits; the green areas are from non-$M_{\rm bc}$ peaking backgrounds or non-$\Sigma_{c}(2455)^{0,++}$ combinatorial backgrounds; the purple dashed curves are the total fitted backgrounds. Here, the $\Sigma_{c}(2455)^{0,++}$ ($M_{\rm bc}$) signal region is required when projecting the corresponding $M_{\rm bc}$ ($M_{\Lambda_{c}^{+}\pi^{\mp}}$) distribution.}
		\end{center}
	\end{figure}

	After releasing the requirements on $M_{\rm bc}$ and the mass window of $\Sigma_{c}(2455)^{0,++}$, the signal yields of $\bar{B}^{0} \to \Sigma_{c}(2455)^{0,++}\pi^{\pm} \bar{p}$ are extracted by unbinned two-dimensional (2D) extended maximum likelihood fits to the $M_{\rm bc}$ and $M_{\Lambda_{c}^{+} \pi^{\mp}}$ distributions of the selected $\bar{B}^{0} \to \Lambda_{c}^{+} \pi^{-} \pi^{+} \bar{p}$ candidates. The 2D fitting function is parameterized as
	\begin{eqnarray*}
		f(M_1,M_2)= N^{\rm sig} s(M_1) S(M_2)+
		N^{\rm bg}_{\rm sb} s(M_1) b'(M_2) \\
		+ N^{\rm bg}_{\rm bs} b(M_1)
		S(M_2)  + N^{\rm bg}_{\rm bb}  g(M_1)
		g'(M_2),
	\end{eqnarray*}
    where $s(M_1)$ and $S(M_2)$ are the 1D signal function in $M_{\rm bc}$ and $M_{\Lambda_c^+\pi^{\mp}}$, respectively, while $b(M_1)$, $g(M_1)$, $b'(M_2)$ and $g'(M_2)$ are the background functions for the same arguments.
	Here, $s(M_1)$ is a Gaussian function, $S(M_2)$ is a non-relativistic Breit-Wigner (BW) function with the phase space factor $p_{\pi^{\mp}}/M_{\Lambda_{c}^{+} \pi^{\mp}}$ considered, convoluted with a triple-Gaussian function whose parameters determined by MC simulation. Moreover, $p_{\pi^{\mp}}$ is the momentum of the selected $\pi^{\mp}$ in the rest frame of $\Lambda_{c}^{+} \pi^{\mp}$ system.
	Here, $b$ and $g$ are ARGUS functions~\cite{argus} while $b'$ and $g'$ are second-order Chebyshev polynomial. All the parameters of the fitting functions are free to float except for those of triple-Gaussian functions.
	The projections of the 2D fits to the selected $\bar{B}^{0} \to \Lambda_{c}^{+}\pi^{-}\pi^{+} \bar{p}$ candidates from data are shown in Fig.~\ref{2dfitB} with the contribution from each component indicated in the legends.
	
	To reduce the influence of the possible intermediate resonances or other non-phase-space contributions in calculating the branching fractions of $\bar{B}^{0}$$\to$  $\Sigma_{c}(2455)^{0,++}$$\pi^{\pm}$$\bar{p}$, the $M_{\Lambda_{c}^{+}\pi^{\mp}\pi^{\pm}}$ versus $M_{\bar{p}\pi^{\pm}}$ planes are divided uniformly into $4\times4$ bins. The $\bar{B}^{0} \to \Sigma_{c}(2455)^{0,++}\pi^{\pm} \bar{p}$ signal yield, $N^{i}_{\Sigma_{c}(2455)^{0,++}}$, where $i$ represents each $M_{\Lambda_{c}^{+}\pi^{\mp}\pi^{\pm}}$ versus $M_{\bar{p}\pi^{\pm}}$ bin, is extracted by the simultaneous fit to all the bins with the same method used in Fig.~\ref{2dfitB}, where the signal functions share the same set of parameters. The total yields of $\bar{B}^{0}\to\Sigma_{c}(2455)^{0} \pi^{+} \bar{p}$ and $\bar{B}^{0}\to\Sigma_{c}(2455)^{++} \pi^{-} \bar{p}$ are $ 767 \pm 44$ and $1213\pm 73$, respectively, obtained by summing the corresponding signal yield in each bin. The total yields are consistent with the overall fit results shown in Fig.~\ref{2dfitB}.
	
	The branching fractions of $\bar{B}^{0} \to \Sigma_{c}(2455)^{0,++}\pi^{\pm} \bar{p}$ are calculated from $ \frac{1}{2\times N_{B\bar{B}} \times \BR(\Upsilon(4S)\to B^{0} \bar{B}^{0})} \times \Sigma_{i}\frac{N^{i}_{\Sigma_{c}(2455)^{0,++}}}{ \varepsilon_{i} }$,
	where, $N_{B\bar{B}} = 772\times10^{6}$ is the number of $B\bar{B}$ pairs and $\BR(\Upsilon(4S)\to B^{0} \bar{B}^{0}) = 0.486\pm0.006$~\cite{PDG}. Furthermore, $\varepsilon_{i}$ $=$ $\Sigma_{j}(\varepsilon^{B}_{j}\times\BR(\Lambda_{c}^{+}\to f_{j}))$ is the reduced detection efficiency in each $M_{\Lambda_{c}^{+}\pi^{\mp}\pi^{\pm}}$ versus $M_{\bar{p}\pi^{\pm}}$ bin, where $f_{j}$ represents $pK^{-}\pi^{+}$, $pK^{0}_{S}$, and $\Lambda\pi^{+}$ for $j = 1,~2,$ and $3$, respectively; $\varepsilon^{B}_{j}$ is the detection efficiency of $\bar{B}^{0} \to \Sigma_{c}(2455)^{0,++}\pi^{\pm} \bar{p}$ with $\Lambda_{c} \to f_{j}$ in the corresponding bin; $\BR(\Lambda_{c}^{+}\to f_{j})$ is the branching fraction of $\Lambda_{c}^{+} \to f_{j}$ including the decay branching fractions of $K_{S}^{0} \to \pi^{+}\pi^{-}$ and $\Lambda \to p\pi^{-}$. Then, the branching fractions of $\bar{B}^{0} \to \Sigma_{c}(2455)^{0,++}\pi^{\pm} \bar{p}$ are calculated to be $\BR(\bar{B}^{0} \to \Sigma_{c}(2455)^{++} \pi^{-} \bar{p}) =(1.84\pm 0.11 )\times10^{-4}$ and $ \BR(\bar{B}^{0} \to \Sigma_{c}(2455)^{0} \pi^{+} \bar{p}) = (1.09 \pm 0.06) \times10^{-4}$. The uncertainties here are statistical only.
	
\begin{figure*}[htbp]
		\begin{center}
			\includegraphics[width=5cm]{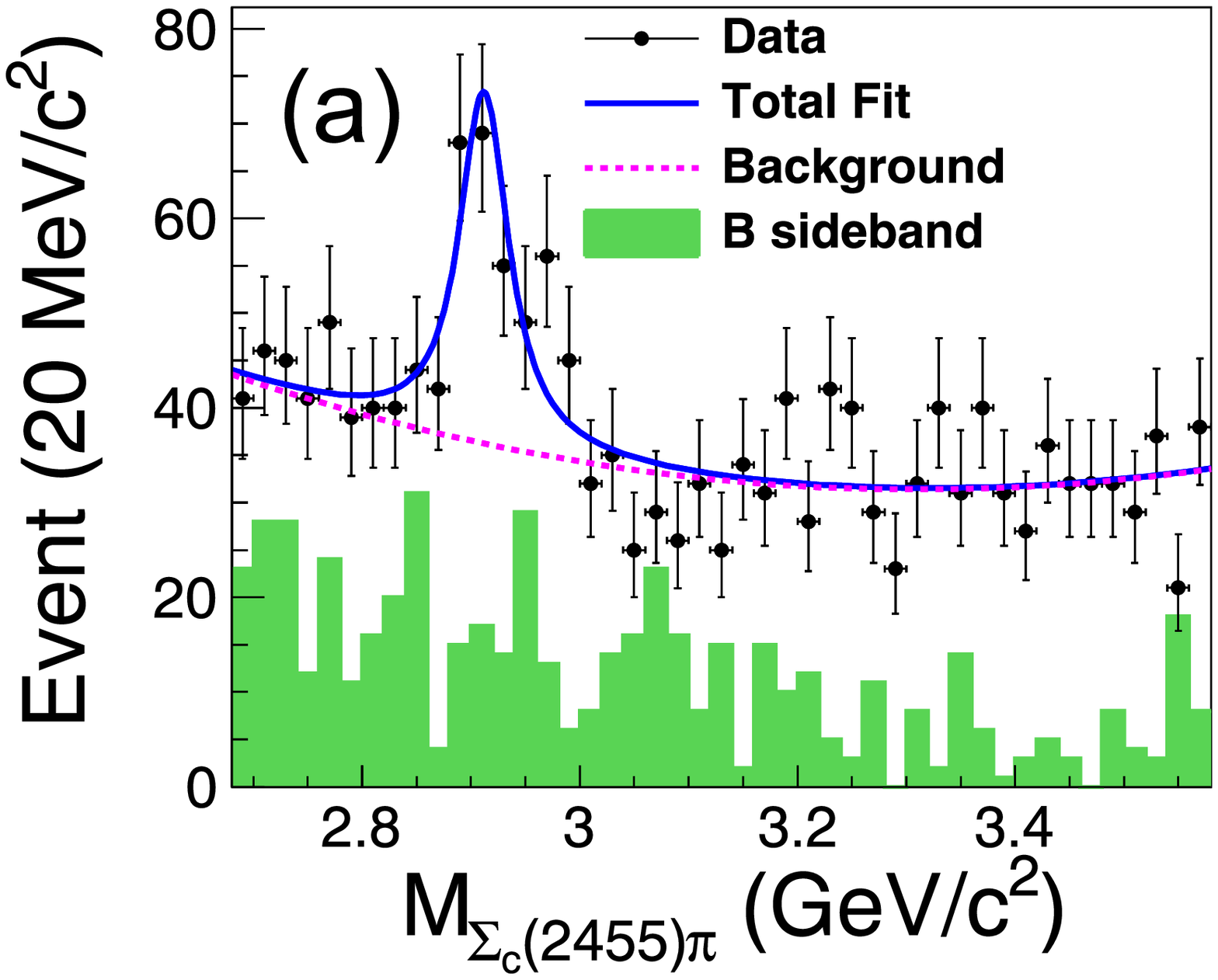}
			\includegraphics[width=5cm]{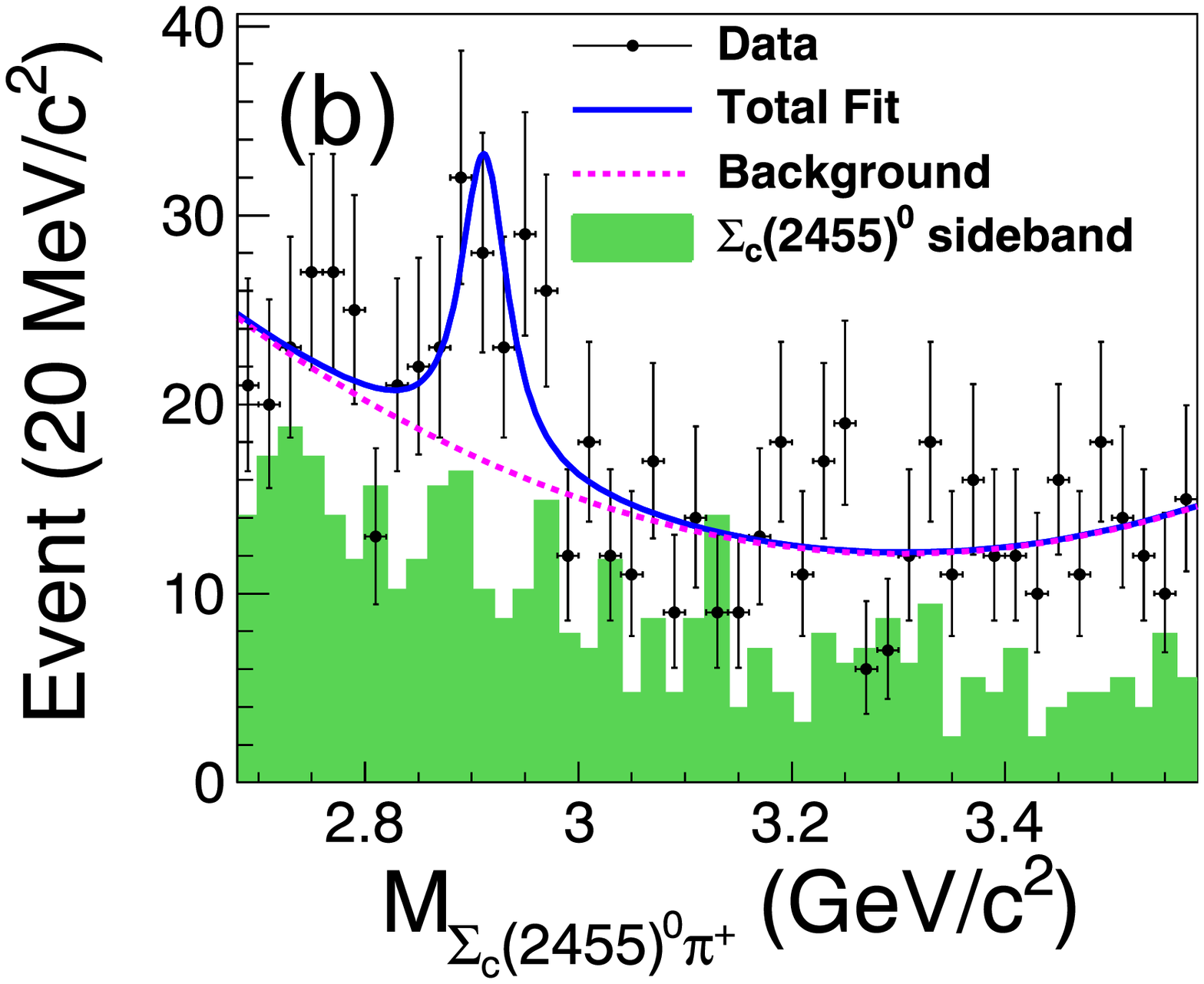}
			\includegraphics[width=5cm]{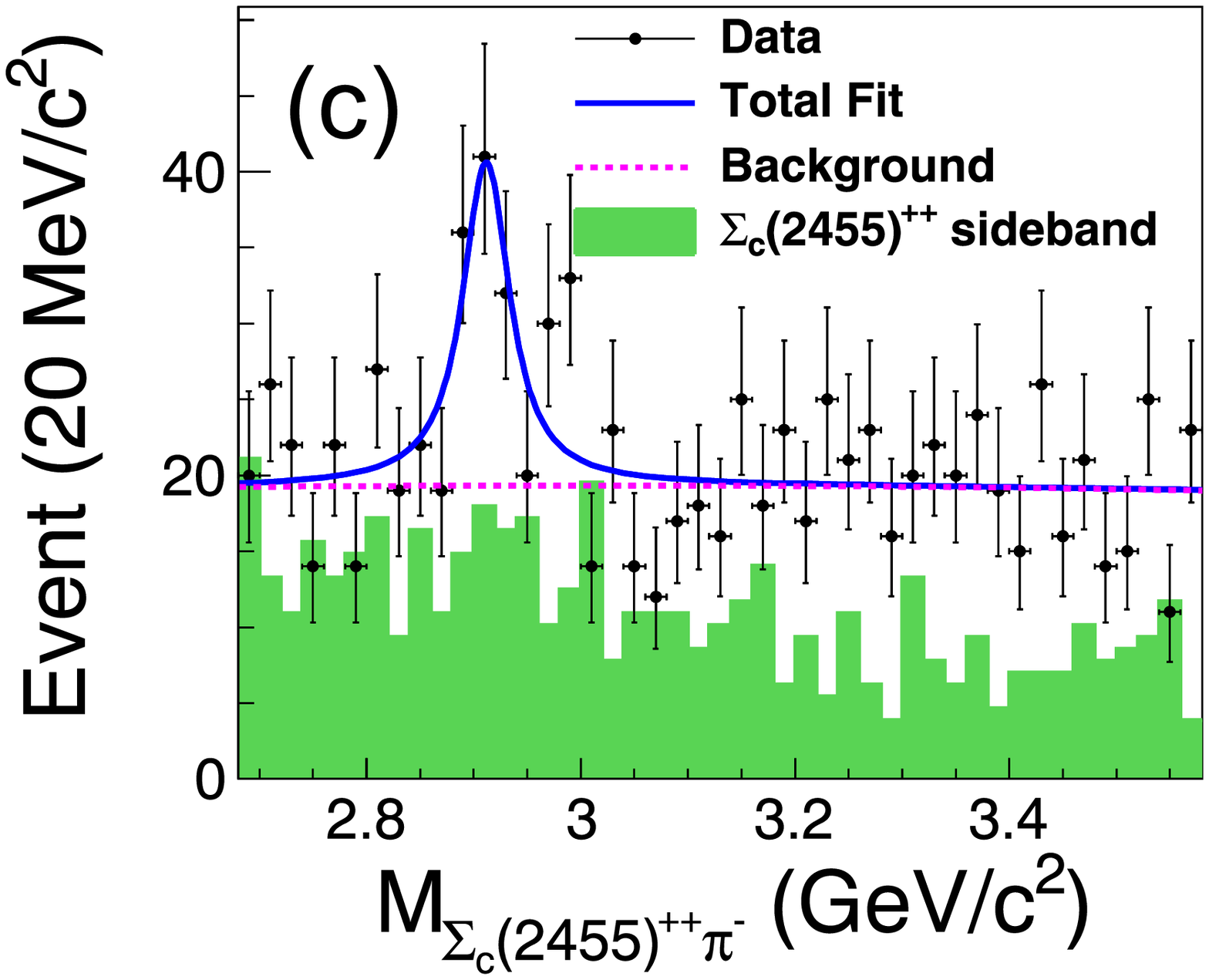}
			\caption{ The fits to the (a) $M_{\Sigma_{c}(2455)\pi}$, (b) $M_{\Sigma_{c}(2455)^{0}\pi^{+}}$, and (c) $M_{\Sigma_{c}(2455)^{++}\pi^{-}}$ distributions of the selected $\bar{B}^{0} \to \Sigma_{c}(2455)^{0,++}\pi^{\pm} \bar{p}$ candidates from data. Here, the data distribution in plot (a) is the sum of those in plots (b) and (c). The dots with error bars represent the data, the solid blue curves are the best fits, and the dashed curves are the fitted backgrounds.}\label{fit_lc}
		\end{center}
	\end{figure*}

	We combine the spectra of $M_{\Sigma_{c}(2455)^{0}\pi^{+}}$ and $M_{\Sigma_{c}(2455)^{++}\pi^{-}}$ (denoted hereinafter as the $M_{\Sigma_c(2455)\pi}$ spectrum) to search for a possible resonance. We estimate the background contributions from non-$M_{\rm bc}$ peaking backgrounds using the events in the three blue sideband regions minus the events in the two red sideband regions in Fig.~\ref{mbcdm}, which are denoted as $B$ sidebands, and the sidebands of $\Sigma_{c}(2455)^{0,++}$, defined as $2.470$ MeV/$c^{2}$ $<M_{\Lambda_{c}^{+}\pi^{\mp}}< 2.491$ MeV/$c^{2}$ or $2.425$ MeV/$c^{2}$ $<M_{\Lambda_{c}^{+}\pi^{\mp}}< 2.437$ MeV/$c^{2}$, to estimate the non-$\Sigma_{c}(2455)^{0,++}$ backgrounds. The distributions of the (a) $M_{\Sigma_{c}(2455)\pi}$, (b) $M_{\Sigma_{c}(2455)^{0}\pi^{+}}$, and (c) $M_{\Sigma_{c}(2455)^{++}\pi^{-}}$ of the selected $\bar{B}^{0} \to \Sigma_{c}(2455)^{0,++} \pi^{\pm} \bar{p}$ candidates in the $\bar{B}^{0}$ signal region and the corresponding $\Sigma_c(2455)$ signal region are shown in Fig.~\ref{fit_lc},
	where a structure around 2.91 GeV/$c^{2}$ can be seen in all plots that cannot be well described by any known resonance. The filled histograms in plots (a), (b), and (c) are from the normalized $B$ sidebands, $\Sigma_{c}(2455)^{0}$ sidebands, and $\Sigma_{c}(2455)^{++}$ sidebands, respectively. There is no peaking contribution from any sideband.

    To determine the parameters of the structure, an unbinned extended maximum likelihood fit is performed to the $M_{\Sigma_{c}(2455)\pi}$ spectrum. The signal shape is a non-relativistic BW convoluted with a Gaussian function (whose width equals to 5.3 MeV/$c^{2}$ determined from MC simulation) with the detection efficiency curve considered. The background is represented with a second-order Chebyshev polynomial. The corresponding fitted signal yield of the structure is  $150 \pm 40$; its mass and width are determined to be $(2913.8 \pm 5.6)$ MeV/$c^{2}$ and $(51.8\pm20.0)$ MeV, respectively. For the mass measurement, the $-$1.5 MeV/$c^{2}$ shift between the output and input mass determined by MC simulation has been corrected (``mass correction factor"). The uncertainties here are statistical only. The statistical significance of the structure is $6.1 \sigma$, estimated from the difference of the logarithmic likelihoods of the fits without and with a signal component with the difference in the number of degrees of freedom, 3, considered~\cite{significance}. Alternative fits to the $M_{\Sigma_{c}(2455)\pi}$ spectrum are performed: (a) using a first or third-order polynomial as the background shape; (b) changing the mass resolution by 10\%; and (c) using an energy-dependent relativistic BW function as the signal shape. The statistical significances of the structure are larger than $5.8\sigma$ in all cases. When only taking the contributions of $\Lambda_{c}(2880)^{+}$ and $\Lambda_{c}(2940)^{+}$ as the signal shapes in the fit with their parameters constrained to be within $1\sigma$ of their world average values~\cite{PDG}, their significances are $1.5\sigma$ and $2.6\sigma$, respectively. However, when introducing $\Lambda_{c}(2880)^{+}$ and $\Lambda_{c}(2940)^{+}$ as additional background components into the above fit with the new structure, their yields are consistent with zero and the significance of the new structure decreases to $4.2\sigma$. Therefore, we take the fit with only one signal component as nominal result, and take $4.2\sigma$ as the signal significance of the new structure with the systematic uncertainty included.
	
	The known particle with the closest mass and width to the structure is $\Lambda_{c}(2940)^{+}$. However, the mass of the structure differs from that of $\Lambda_{c}(2940)^{+}$\cite{PDG} by $3.8\sigma$ with systematic uncertainty described below considered.
	Since the mass difference between the structure and $\Lambda_{c}(2940)^{+}$ agrees with the expected mass gap between $\Lambda_{c}(\frac{1}{2}^{-},2P)$ and $\Lambda_{c}(\frac{3}{2}^{-},2P)$ state in quark models~\cite{Capstick:1986ter,Ebert2,Chen1,bing}, the structure is a good candidate for the $\Lambda_{c}(\frac{1}{2}^{-},2P)$ state and is tentatively named as $\Lambda_{c}(2910)^{+}$. The $\bar{B}^{0} \to \Lambda_{c}(2910)^{+}\bar{p}$ and $\Lambda_{c}(2910)^{+} \to \Sigma_{c}(2455)^{0,++}\pi^{\pm}$ are $S$-wave decays under this assumption. However, further study is needed to confirm whether this state is an excited $\Lambda_{c}$ or $\Sigma_{c}$.

	To determine the signal yields of $\bar{B}^{0} \to \Lambda_{c}(2910)^{+}\bar{p}$ with $\Lambda_{c}(2910)^{+} \to \Sigma_{c}(2455)^{0,++}\pi^{\pm}$, a simultaneous unbinned extended maximum-likelihood fit to the $M_{\Sigma_{c}(2455)^{0}\pi^{+}}$ and $M_{\Sigma_{c}(2455)^{++}\pi^{-}}$ spectra is performed, where the signal function is the same to both spectra. The fit functions are the same as those used in the nominal fit to $M_{\Sigma_{c}(2455)\pi}$ spectrum above with all the parameters free to float. The fit results are shown in panels (b) and (c) of Fig.~\ref{fit_lc}. The fitted signal yields are $N_{\Sigma_{c}(2455)^{0}\pi^{+}} = 63\pm24$ and $N_{\Sigma_{c}(2455)^{++}\pi^{-}} = 83 \pm 23$ for $\Lambda_{c}(2910)^{+} \to \Sigma_{c}(2455)^{0}\pi^{+}$ and $\Lambda_{c}(2910)^{+} \to \Sigma_{c}(2455)^{++}\pi^{-}$, respectively. The fitted mass and width of $\Lambda_{c}(2910)^{+}$ are $(2914.7\pm 5.6)$ MeV/$c^{2}$ and $(50.1 \pm 20.5)$ MeV, respectively, which are consistent with those from the fit to the $M_{\Sigma_{c}(2455)\pi}$ spectrum.
	
	The branching fraction product of $\BR(\bar{B}^{0} \to \Lambda_{c}(2910)^{+}\bar{p})\times\BR(\Lambda_{c}(2910)^{+} \to \Sigma_{c}(2455)^{0,++}\pi^{\pm}$) is calculated with
	$
	N_{\Sigma_{c}(2455)^{0,++}\pi^{\pm}}/(2\times N_{B\bar{B}} \times \BR(\Upsilon(4S)\to B^{0} \bar{B}^{0}) \times \Sigma_{i}(\BR(\Lambda_{c}^{+}\to f_{i}) \times \varepsilon_{i}^{\Lambda_{c}(2910)^{+}})),
	$
	where $\varepsilon_{i}^{\Lambda_{c}(2910)^{+}}$ is the detection efficiency of $\bar{B}^{0} \to \Lambda_{c}(2910)^{+}\bar{p}$ with $\Lambda_{c}(2910)^{+} \to \Sigma_{c}(2455)^{0,++}\pi^{\pm}$, $\Sigma_{c}(2455)^{0,++}\to \Lambda_{c}^{+}\pi^{\pm}$, and $\Lambda_{c}^{+}\to f_{i}$, which is $10.60\%$, $12.14\%$, and $11.24\%$ for $\Lambda_{c}^{+}\to pK^{-}\pi^{+}$, $pK_{S}^{0}$, and $\Lambda\pi^{+}$, respectively. The detection efficiencies here include the particle identification (PID) correction factors described below and the decay branching fractions of $K_{S}^{0} \to \pi^{+}\pi^{-}$ and $\Lambda \to p\pi^{-}$. The detection efficiencies are the same for $\Sigma_{c}(2455)^{++}$ and $\Sigma_{c}(2455)^{0}$ intermediate states, according to the MC simulations. The branching fraction products are calculated to be $\BR(\bar{B}^{0} \to \Lambda_{c}(2910)^{+}\bar{p})\times\BR(\Lambda_{c}(2910)^{+} \to \Sigma_{c}(2455)^{0}\pi^{+}) = (9.5 \pm 3.6)\times10^{-6}$ and $\BR(\bar{B}^{0} \to \Lambda_{c}(2910)^{+}\bar{p})\times\BR(\Lambda_{c}(2910)^{+} \to \Sigma_{c}(2455)^{++}\pi^{-}) = (1.24 \pm 0.35)\times10^{-5}.$ The errors here are statistical only.
	
	There are several sources of systematic uncertainties in the
	branching fraction measurements. Using $D^{*+} \to D^{0} \pi^{+}$, $D^{0} \to K^{-} \pi^{+}$, and $\Lambda \to p \pi^{-}$ samples, the efficiency ratios between data and MC simulations are $0.998\pm 0.013$, $0.970\pm 0.006$, $0.900 \pm 0.005$, and $0.987\pm 0.005$ for kaon, pion, proton from $\Lambda_{c}^{+}$, and proton directly from $\bar{B}^{0}$, respectively, whose central values are taken as the efficiency correction factors and the errors are taken as the systematic uncertainties due to PID.
	The uncertainties on the branching fractions of $\Lambda_{c}^{+}$ decay chains are $5.1\%$, $5.1\%$, and $5.4\%$ for $\Lambda_{c}^{+} \to pK^{-}\pi^{-}$, $pK_{S}^{0}$, and $\Lambda\pi^{+}$ modes~\cite{PDG}, respectively.
	The uncertainties on the detection efficiency include those from PID, the branching fractions of $\Lambda_{c}^{+}$ decays, tracking efficiency (0.35\%/track), as well as $\Lambda$ (2.95\%) and $K_{S}^{0}$ (0.5\%) selection efficiencies. Assuming all the sources of the above systematic uncertainties are independent, the uncertainties from the same sources are summed linearly weighted by the expected signal yields over the three $\Lambda_c^{+}$ decay modes. Then, the uncertainties from different sources are added in quadrature to yield the total uncertainties on
	detection efficiency, which are listed in Table~\ref{tab:err1}.
	
	We estimate the systematic uncertainties on the fitting procedure by changing the order of the background polynomial, the range of the fit, and the mass resolution (enlarged by 10\%). The deviations from the nominal fitted results are taken as systematic uncertainties. For $\BR (\bar{B}^{0}$ $\to$ $\Sigma_{c}(2455)^{0,++}\pi^{\pm} \bar{p})$, the fitting uncertainties in $M_{\Lambda_{c}^{+}\pi^{\mp}\pi^{\pm}}$ versus $M_{\bar{p}\pi^{\pm}}$ bins are summed in quadrature weighted by $1/\varepsilon_{i}$. These uncertainties are added in quadrature to yield the total uncertainties due to fit.
	
	The uncertainties on the world average value of $\BR(\Upsilon(4S)\to B^0\bar{B}^0)$ and $N_{\Upsilon(4S)}$ are 1.2\% and 1.37\%, respectively. Thus, the uncertainty of the $\bar{B}^{0}$ count is $1.8\%$.
	
	Assuming all sources listed in Table~\ref{tab:err1} are independent, the uncertainties from different sources are added in quadrature to yield the total systematic uncertainties.
	
	\begin{table}[t]
		\caption{\label{tab:err1}  Summary of the systematic uncertainties on the branching fraction measurements (\%). Here, $\BR_{i}$ means $\BR (\bar{B}^{0} \to \Sigma_{c}(2455)^{++}\pi^{-} \bar{p})$, $\BR (\bar{B}^{0} \to \Sigma_{c}(2455)^{0}\pi^{+} \bar{p})$, $\BR(\bar{B}^{0} \to \Lambda_{c}(2910)^{+}\bar{p})\times\BR(\Lambda_{c}(2910)^{+} \to \Sigma_{c}(2455)^{0}\pi^{+})$, and $\BR(\bar{B}^{0} \to \Lambda_{c}(2910)^{+}\bar{p})\times\BR(\Lambda_{c}(2910)^{+} \to \Sigma_{c}(2455)^{++}\pi^{-})$ for $i = 1,~2,~3,$ and $4$, respectively.}
		\renewcommand\arraystretch{1.2}
			\setlength{\tabcolsep}{3.mm}{
				\begin{tabular}{cccccp{3cm}}
					\hline\hline
					$\BR_{i}$ & Detection efficiency  &  Fit  &  $N_{B}$ & Sum \\
					\hline
					$\BR_{1}$ & 5.9  & 1.6  & 1.8  & 6.4 \\
					$\BR_{2}$  & 5.9  & 1.9  & 1.8  & 6.5 \\
					$\BR_{3}$  & 5.6  & 16  & 1.8  & 17\\
					$\BR_{4}$  & 5.8  & 5.6  & 1.8  &  8.2 \\
					\hline\hline
				\end{tabular}
			}
		\end{table}
		
		The following systematic uncertainties are considered for the mass and width of $\Lambda_{c}(2910)^{+}$. Half of the mass correction factor is taken as a systematic uncertainty. By changing the order of the background polynomial and fit region, the differences in the fitted $\Lambda_{c}(2910)^{+}$ mass (3.42 MeV/$c^2$) and width (18.3 MeV) are taken as systematic uncertainties.
		By replacing the non-relativistic BW function by a relativistic BW function with a mass-dependent width of $\Gamma_{t} = \Gamma^{0}_{t}
		\frac{\Phi(M_{\Sigma_{c}\pi})}{\Phi(M_{\Lambda_c(2910)^{+}})}$, where
		$\Gamma_{t}^{0}$ is the width of the resonance, $\Phi(M_{x}) = \frac{P}{M_{x}}$ is the $S$-wave phase space factor ($P$ is the $\pi$
		momentum in the $\Sigma_c\pi$ or $\Lambda_{c}(2910)^{+}$ center-of-mass frame), the difference in the measured $\Lambda_{c}(2910)^{+}$ mass (1.2 MeV/$c^2$) is taken as a systematic uncertainty. When considering the background contributions from $\Lambda_{c}(2880)^{+}$ and $\Lambda_{c}(2940)^{+}$, by changing their masses and widths by $\pm1\sigma$~\cite{PDG}, the differences in mass and width of $\Lambda_{c}(2910)^{+}$ are 1.0 MeV/$c^2$ and 4.3 MeV, respectively, which are taken as systematic uncertainties.
		Assuming all the sources are independent, we add them in quadrature to
		obtain the total systematic uncertainties on the $\Lambda_{c}(2910)^{+}$
		mass and width, which are $3.8$ MeV/$c^2$ and $18.8$ MeV, respectively.


		In summary, based on $772\times10^{6}$ pairs of $B\bar{B}$ data samples collected with the Belle detector at the KEKB asymmetric-energy $e^+e^-$ collider, we analyze the $\bar{B}^{0} \to \Sigma_{c}(2455)^{++,0} \pi^{\mp} \bar{p}$ decays with the branching fractions measured to be $\BR(\bar{B}^{0} \to \Sigma_{c}(2455)^{++} \pi^{-} \bar{p}) = (1.84\pm 0.11 \pm 0.12)\times10^{-4}$ and $ \BR(\bar{B}^{0} \to \Sigma_{c}(2455)^{0} \pi^{+} \bar{p}) = (1.09 \pm 0.06 \pm 0.07) \times10^{-4}$, which are consistent with the previous measurements~\cite{PDG,cleo_old,belle_old,belle_old2,babar_old} with improved precision. A structure around $2.91$ GeV/$c^{2}$ is found in the $M_{\Sigma_{c}(2455)\pi}$ spectrum with a statistical significance of $6.1 \sigma$. The significance changes to 4.2$\sigma$ when introducing possible background contributions from $\Lambda_{c}(2880)^{+}$ and $\Lambda_{c}(2940)^{+}$. The mass and width of the state are measured to be $(2913.8 \pm 5.6 \pm 3.8)$ MeV/$c^{2}$ and $(51.8\pm20.0 \pm 18.8)$ MeV, respectively. This state is possibly a good candidate for $\Lambda_{c}(\frac{1}{2}^{-},2P)$ and is tentatively named as $\Lambda_{c}(2910)^{+}$, with its nature needing more investigation. The products of branching fractions concerning the $\Lambda_{c}(2910)^{+}$ are measured to be $\BR(\bar{B}^{0} \to \Lambda_{c}(2910)^{+}\bar{p})\times\BR(\Lambda_{c}(2910)^{+} \to \Sigma_{c}(2455)^{++}\pi^{-}) = (1.24 \pm 0.35 \pm 0.10)\times10^{-5}$, and $\BR(\bar{B}^{0} \to \Lambda_{c}(2910)^{+}\bar{p})\times\BR(\Lambda_{c}(2910)^{+} \to \Sigma_{c}(2455)^{0}\pi^{+}) = (9.5 \pm 3.6 \pm 1.6)\times10^{-6}$. Here, the first and second uncertainties are statistical and systematic, respectively. The $\BR(\bar{B}^{0} \to \Sigma_{c}(2455)^{0,++} \pi^{\pm} \bar{p})$ measurements in this analysis supersede the previous Belle measurements~\cite{belle_old}.
		

This work, based on data collected using the Belle detector, which was
operated until June 2010, was supported by 
the Ministry of Education, Culture, Sports, Science, and
Technology (MEXT) of Japan, the Japan Society for the 
Promotion of Science (JSPS), and the Tau-Lepton Physics 
Research Center of Nagoya University; 
the Australian Research Council including grants
DP180102629, 
DP170102389, 
DP170102204, 
DE220100462, 
DP150103061, 
FT130100303; 
Austrian Federal Ministry of Education, Science and Research (FWF) and
FWF Austrian Science Fund No.~P~31361-N36;
the National Natural Science Foundation of China under Contracts
No.~11675166,  
No.~11705209;  
No.~11975076;  
No.~12135005;  
No.~12175041;  
No.~12161141008; 
Key Research Program of Frontier Sciences, Chinese Academy of Sciences (CAS), Grant No.~QYZDJ-SSW-SLH011; 
Project ZR2022JQ02 supported by Shandong Provincial Natural Science Foundation;
the Ministry of Education, Youth and Sports of the Czech
Republic under Contract No.~LTT17020;
the Czech Science Foundation Grant No. 22-18469S;
Horizon 2020 ERC Advanced Grant No.~884719 and ERC Starting Grant No.~947006 ``InterLeptons'' (European Union);
the Carl Zeiss Foundation, the Deutsche Forschungsgemeinschaft, the
Excellence Cluster Universe, and the VolkswagenStiftung;
the Department of Atomic Energy (Project Identification No. RTI 4002) and the Department of Science and Technology of India; 
the Istituto Nazionale di Fisica Nucleare of Italy; 
National Research Foundation (NRF) of Korea Grant
Nos.~2016R1\-D1A1B\-02012900, 2018R1\-A2B\-3003643,
2018R1\-A6A1A\-06024970, RS\-2022\-00197659,
2019R1\-I1A3A\-01058933, 2021R1\-A6A1A\-03043957,
2021R1\-F1A\-1060423, 2021R1\-F1A\-1064008, 2022R1\-A2C\-1003993;
Radiation Science Research Institute, Foreign Large-size Research Facility Application Supporting project, the Global Science Experimental Data Hub Center of the Korea Institute of Science and Technology Information and KREONET/GLORIAD;
the Polish Ministry of Science and Higher Education and 
the National Science Center;
the Ministry of Science and Higher Education of the Russian Federation, Agreement 14.W03.31.0026, 
and the HSE University Basic Research Program, Moscow; 
University of Tabuk research grants
S-1440-0321, S-0256-1438, and S-0280-1439 (Saudi Arabia);
the Slovenian Research Agency Grant Nos. J1-9124 and P1-0135;
Ikerbasque, Basque Foundation for Science, Spain;
the Swiss National Science Foundation; 
the Ministry of Education and the Ministry of Science and Technology of Taiwan;
and the United States Department of Energy and the National Science Foundation.
These acknowledgements are not to be interpreted as an endorsement of any
statement made by any of our institutes, funding agencies, governments, or
their representatives.
We thank the KEKB group for the excellent operation of the
accelerator; the KEK cryogenics group for the efficient
operation of the solenoid; and the KEK computer group and the Pacific Northwest National
Laboratory (PNNL) Environmental Molecular Sciences Laboratory (EMSL)
computing group for strong computing support; and the National
Institute of Informatics, and Science Information NETwork 6 (SINET6) for
valuable network support.


\begin{thebibliography}{99}
\bibitem{charge-conjugate} Inclusion of charge conjugate states is implicit unless otherwise stated.
\end{thebibliography}
	\end{document}